\documentclass[5p, twocolumn, floatfix]{elsarticle}
\usepackage{amsmath, amsthm, amssymb}
\usepackage{bbm}
\usepackage{epsfig}
\usepackage{slashed}
\usepackage{color}
\usepackage{feynmf}
\graphicspath{{graphics/}}
\usepackage{hyperref}

\renewcommand{\=}{\;=\;}

\newcommand{\eq}[1]{Eq.~(\ref{#1})}
\newcommand{\eqs}[1]{Eqs.~(\ref{#1})}
\newcommand{\fig}[1]{Fig.~\ref{#1}}

\newcommand{\Tr}{\mbox{Tr}}

\newcommand{\SI}{S^{-1}}

\newcommand{\va}[1]{\vert\vec #1 \vert}

\newcommand{\vv}{{\cal V}}
\newcommand{\one}{\mathbbm{1}}

\newcommand{\bit}{\begin{itemize}}
\newcommand{\eit}{\end{itemize}}
\newcommand{\beq}{\begin{equation}}
\newcommand{\eeq}{\end{equation}}
\newcommand{\bea}{\begin{align}}
\newcommand{\eea}{\end{align}\]}
\newcommand{\lb}{\left(}
\newcommand{\rb}{\right)}
\def\bseq#1\eseq{\begin{equation}\begin{split}#1\end{split}\end{equation}}

\begin{document}

 \title{Dyson-Schwinger study of chiral density waves in QCD}
 \date{\today}
 \author[tud]{D. M\"uller} \author[tud]{M. Buballa} \author[tud,gsi]{J. Wambach}
\address[tud]{Institut f\"ur Kernphysik (Theoriezentrum), Technische Universit\"at Darmstadt, Germany}
\address[gsi]{GSI Helmholtzzentrum f\"ur Schwerionenforschung, Darmstadt, Germany}

\begin{abstract}
The formation of inhomogeneous chiral condensates in QCD matter at nonzero density 
and temperature is investigated for the first time with Dyson-Schwinger equations. 
We consider two massless quark flavors in a so-called chiral density wave, where scalar and pseudoscalar quark condensates 
vary sinusoidally along one spatial dimension. 
We find that the inhomogeneous region covers the major part of the spinodal region of
the first-order phase transition which is present when the analysis is restricted to 
homogeneous phases.
The triple point where the inhomogeneous phase meets the homogeneous phases with broken 
and restored chiral symmetry, respectively, coincides, within numerical accuracy,
with the critical point of the homogeneous calculation. 
At zero temperature, the inhomogeneous phase seems to extend to arbitrarily high
chemical potentials, as long as pairing effects are not taken into account.
\end{abstract}

\maketitle

The properties of strong-interaction matter under extreme conditions, such as high temperature
or density, and the corresponding phase structure of Quantum Chromodynamics (QCD) 
are subject of extensive theoretical and experimental 
investigations~\cite{BraunMunzinger:2009zz,Fukushima:2010bq}. 
At vanishing net baryon density, first-principle lattice gauge calculations have revealed
that the approximate chiral symmetry of QCD, which is spontaneously broken
at low temperature, gets restored at high temperature in a cross-over transition~\cite{Borsanyi:2010bp,Bazavov:2011nk}. 
At low temperature and high baryon density, where lattice calculations  are inhibited
by the sign problem, effective model studies typically predict that chiral symmetry is
restored in a first-order phase transition, which weakens with increasing temperature
and eventually ends at a critical point~\cite{Asakawa:1989bq}.
More recently, this picture was confirmed with 
Dyson-Schwinger equations  (DSEs) 
applied to QCD~\cite{Fischer:2009gk,Fischer:2010fx,Fischer:2011mz,Fischer:2012vc}. 

A basic assumption in these investigations was that the phases are homogeneous, i.e.,
in particular, the chiral order parameter is constant in space. 
On the other hand, phases with non-uniform chiral order parameters have been proposed
already long time ago, see Ref.~\cite{Broniowski:2011ef} for a brief historical review.
Starting with Migdal's $p$-wave pion condensation~\cite{Migdal:1973zm},
the idea was generalized to relativistic systems~\cite{Dautry:1979bk,Broniowski:1990dy,Nakano:2004cd}
and studied in high-density QCD with large number of colors, applying weak-coupling 
methods~\cite{Deryagin:1992rw,Shuster:1999tn}.
More recently, it gained new attention after it was found in effective models 
that the first-order chiral phase boundary between homogeneous phases
is covered completely by an inhomogeneous phase~\cite{Nickel:2009ke,Nickel:2009wj}.

The aim of our work is to study the chiral phase transition with the possibility of inhomogeneous condensates in strong-coupling QCD with DSEs. 
We restrict ourselves to a so-called chiral density wave (CDW),  
where the chiral condensate rotates along the chiral circle, when moving into a fixed direction.
Specifically, the scalar and pseudoscalar condensates behave like
\beq
\label{eq:cond}
\langle \bar q q \rangle \propto \cos (Qz)\,,\quad
\langle \bar q i\gamma_5 \tau_3 q \rangle \propto \sin (Qz)\,,
\eeq
with $Q$ being the modulus of a wave vector, which we have chosen to point into the
$z$ direction. 
Moreover, as indicated by the Pauli matrix $\tau_3$,
we have chosen the third isospin component of the pseudoscalar condensate.
Assuming isospin invariance of the QCD Lagrangian,
this can be done without loss of generality.

\begin{figure}
	\centering
		\includegraphics[scale=0.9]{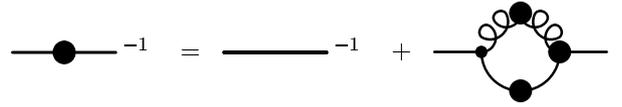}	
\caption{Dyson-Schwinger equation for the full quark propagator. Plain lines represent quark propagators, the curly line the gluon propagator. Thick dots represent dressed quantities.}
\label{feyn_qdse}
\end{figure}

The DSE for the dressed quark propagator $S$ is diagrammatically depicted in 
Fig.~\ref{feyn_qdse}. In coordinate space with Euclidean metric, it is given by 
$S^{-1}(x,x') = Z_2\lb S_0^{-1}(x,x') + \Sigma(x,x')\rb$,
depending on two space-time variables, $x$ and $x'$.
$S_0$ denotes the bare propagator, $\Sigma$ the selfenergy and $Z_2$ is the
wave-function renormalization constant of the quark field. 

In homogeneous, i.e., translationally invariant matter, the propagator depends only on the 
relative coordinate $x-x'$. In momentum space, this translates into a dependence on a single
4-momentum, and the inverse propagator at temperature $T$ and chemical potential $\mu$
can be parametrized as
\beq
S_\mathit{hom}^{-1}(p) = -i\omega_n\gamma_4 C(p)-i{\vec\gamma}\cdot {\vec p}A(p) + B(p)\,,
\label{eq:Shom}
\eeq
with $p:=(\vec p, p_4=\omega_n+i\mu)$, the Matsubara frequencies $\omega_n=(2n+1)\pi T$,
and three dressing functions $A$, $B$, and $C$.
In vacuum, due to Lorentz covariance, these functions depend on $p^2$ only and $A(p) = C(p)$. 
The wave-function renormalization constant $Z_2$ is then fixed by the condition that 
$A(p)|_{p^2=\nu^2} =1$ at an arbitrary renormalization point $\nu$.
This prescription remains valid for our analysis of inhomogeneous phases
since $Z_2$ is always fixed in vacuum, which is homogeneous.

In inhomogeneous matter, the quark selfenergy and, hence, the propagator
depend separately on both coordinates $x$ and $x'$, or, equivalently, on the relative coordinate
$x-x'$ and the center of momentum coordinate $(x+x')/2$.
In momentum space they thus depend on two momenta, and the DSE reads
\beq
\label{eq:qdse}
S^{-1}(p,p') = Z_2\lb S_0^{-1}(p,p') + \Sigma(p,p')\rb .
\eeq
Here $p$ and $p'$ correspond to the out- and ingoing momenta of the quark,
which do not need to be identical. 
Physically, this means that the inhomogeneous condensates carry momentum,
so that the quark can change its momentum by scattering off the condensate. 

The (inverse) propagator can be viewed as a continuous matrix in momentum space.
For the formal manipulations to be discussed below it is useful to 
introduce a finite quantization volume $V$ in 3-space and take periodic boundary 
conditions. Thus, together with the finite extent in the imaginary time direction, 
the 4-volume is finite as well, $\vv = V \times [0,1/T]$, and the 4-momenta take discrete 
values.
At the end we will take the limit $V\rightarrow \infty$, so that 
the 3-momenta will be continuous variables again.
Momentum sums and Kronecker symbols should then be replaced as
\beq
\frac{1}{\vv}\sum\limits_p \rightarrow T\sum_{n} \int\frac{d^3p}{(2\pi)^3}\,,
\quad
\vv\delta_{p,p'} \rightarrow \frac{1}{T} \,\delta_{n,n'}(2\pi)^3\delta(\vec p- \vec p\,'),
\label{eq:infV}
\eeq
where $n$ and $n'$ label the Matsubara frequencies.

In this Letter we consider two quark flavors with vanishing bare mass. 
Since the bare propagator is constructed on the (homogeneous) perturbative ground 
state, it stays diagonal in momentum space and keeps its familiar form,
\beq
S_0^{-1}(p,p')= - i\slashed p\,\vv\delta_{p,p'}\,.
\eeq
The quark selfenergy is given by (see Fig.~\ref{feyn_qdse})
\beq
\label{eq:qsig}
Z_2\Sigma(p,p') \= 
g^2 \frac{1}{\vv}\sum\limits_{q}\Gamma_\mu^{a,0}\ S(q,q') D_{\mu\nu}(k)\Gamma^a_\nu(q',p'),
\eeq
with the QCD coupling constant $g$,
the bare and dressed quark-gluon vertices, $g\Gamma_\mu^{a,0}$ and 
$g\Gamma^a_\nu$, respectively, 
and the dressed gluon propagator $D_{\mu\nu}$.
As detailed below, we neglect possible modifications of 
these quantities with respect to the homogeneous case. 
As a consequence, the gluon propagator depends only on a single momentum variable, 
and 4-momentum is conserved at the vertices, i.e., $k=p-q$ and $q'=q+p'-p$. 

The bare vertex is given by $\Gamma_\mu^{a,0} = Z_{1F}\gamma_\mu\lambda^a/2$
with a renormalization constant $Z_{1F}$ and the Gell-Mann matrix $\lambda^a$.
The dressed gluon propagator and the dressed vertex are in principle given by their
own DSEs. Since these depend on even higher $n$-point functions, truncations are 
necessary to get a closed set of equations.  
Here we adopt the truncation scheme described in Ref.~\cite{Muller:2013pya},
and we refer to that reference for details and parameters.

In this scheme the dressed vertex is taken to have the same structure as the bare vertex,
$\Gamma^a_\mu(p,q) = \Gamma(p-q) \gamma_\mu\lambda^a/2$
with a dressing function  $\Gamma(k)$, 
which has the correct perturbative running in the ultraviolet and a phenomenological
enhancement in the infrared. 
The gluon propagator is based on a parametrization of lattice data for the Yang-Mills system,
which is corrected for quark effects by perturbatively adding a polarization loop in 
hard-thermal-loop--hard-dense-loop approximation. 
This accounts for Debye screening and Landau damping at high temperature or chemical 
potential, but neglects the dressing of the quarks in the polarization loops. 
As a consequence, the dressed gluon propagator remains diagonal in momentum space,
as already mentioned above. 

The task is now to generalize the structure \eq{eq:Shom}
of the quark propagator in a homogeneous ground state
to an inhomogeneous medium where the quark condensate takes the form 
of a CDW, \eq{eq:cond}.
Starting from the definition $Z_2 S(x,x') = \langle{\cal T} (q(x) \bar q(x'))\rangle$
of the Euclidean propagator, where ${\cal T}$ is denotes the imaginary time ordering operator,
the condensates are related to the propagator as
\beq
\langle \bar q {\cal O} q \rangle = 
-Z_2 \frac{1}{\vv^2}\sum\limits_{p,p'}e^{i (p-p')\cdot x} ~ \Tr\left[ {\cal O} S(p, p')\right ],
\eeq
with the trace in Dirac, flavor and color space.
Comparing this with \eq{eq:cond}, we find that the desired spatial behavior is obtained if
\beq
\Tr\left[ (\one\pm\gamma_5\tau_3) S(p, p')\right ]  \propto \delta_{p,p'\mp Q}\,
 \eeq
 with  the wave vector $Q\equiv Q e_3$ being a 4-vector of length $Q$, pointing to the 
 3-direction.
This suggests to generalize the dressing function $B$ , which is the only 
chiral-symmetry breaking term in \eq{eq:Shom}, in a similar way.
Specifically, we make the ansatz that the inverse propagator $S^{-1}(p,p')$ contains a term
\beq
\label{eq:binh}
B(p,p') = \frac{B(p) + B(p')}{2} \sum\limits_{s=\pm}
                \frac{\one+s\gamma_5\tau_3}{2} \,\vv\delta_{p,p'-sQ}
\eeq
where $B(p)$ is closely related to the $B$ function of the homogeneous case.
In fact,  when we take $Q=0$, the matrix $B(p,p')$ becomes purely scalar and
diagonal in momentum space, with the diagonal matrix elements essentially given by 
$B(p)$.\footnote{Apart from a trivial factor of $\vv$ due to the fact that we performed 
a Fourier transform with respect to two space-time arguments instead of only one.}
It is also instructive to compare our ansatz with the Nambu--Jona-Lasinio model where the 
quark selfenergy is local in coordinate space.
For homogeneous matter, this leads to a constant selfenergy in momentum space,
while for inhomogeneous matter the selfenergy can only depend on the difference $p-p'$ 
but not on the sum. For a CDW the selfenergy is then given by \eq{eq:binh} with 
$B(p) = \mathit{const}$.

In our case, $B(p)$ is an unknown function, which must be determined through the DSE.  
To that end, the inverse propagator with the dressing function \eq{eq:binh} must be 
inverted and inserted into \eq{eq:qsig}. It turns out that this induces further structures,
and we need additional dressing functions to achieve a self-consistent solution. 
For instance, since the wave vector $Q = Q e_3$ defines a preferred direction, 
the $A$ function in \eq{eq:Shom} must be replaced by two independent functions, 
corresponding to the momentum component $p_3$ in $Q$ direction and to the perpendicular
part $\vec p_\bot \equiv p_1 e_1 + p_2 e_2$ .
The complete ansatz contains in total 10 dressing functions and reads
\bseq
\label{eq:SIQ}
&\SI(p,p') = -i\Big\lbrace\left[C(p)+\gamma_5 C_5(p)\right](\omega_n+i\mu)\gamma_4
\\
& + \left[E(p)+\gamma_5 E_5(p)\right]p_3\gamma_3  
   + \left[A(p)+\gamma_5 A_5(p)\right] {\vec\gamma \cdot \vec{p}}_\bot \Big\rbrace\vv\delta_{p,p'}
   \\
&+\sum\limits_{s=\pm}\bigg( \bar B(p,p')-i s\gamma_4\gamma_3 \bar F(p,p') 
-i s\gamma_4 \frac{\vec\gamma\cdot{\vec{p}}_\bot}{\va {p_\bot}} \bar G(p,p')
\\
 &~~~-i s\gamma_3\frac{\vec\gamma\cdot{\vec{p}}_\bot}{\va {p_\bot}} \bar H(p,p') \bigg) 
 \frac{(\mathbbm{1}+s\gamma_5\tau_3)}{2}\vv\delta_{p,p'- sQ} 
\eseq
with $\bar B(p,p') = \frac{1}{2}\lb B(p) + B(p')\rb$ and similar for $\bar F$, $\bar G$ and $\bar H$. 

For a general inhomogeneous ansatz, the inversion of $\SI$, which is an infinite matrix in 
momentum space, is a highly non-trivial task.
However, for the CDW it turns out that $\SI$ has a relatively simple block structure
so that it can be inverted analytically. 
The resulting dressed propagator has the same tensor structure as $\SI$.

\begin{figure}
	\centering
		\includegraphics[scale=0.9]{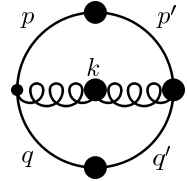}	
\caption{Feynman diagram for the interaction term $\Gamma_2$ of the effective action. The gluon momentum is defined by $k=p-q$ and momentum conservation implies $q'=q+p'-p$.}
\label{feyn_gamma2}
\end{figure}

Starting from \eq{eq:SIQ} one can find self-consistent solutions of the DSE for arbitrary values 
of the wave number $Q$. We thus need an additional constraint to fix $Q$.
This is provided
by the requirement that the free energy of the system is minimal for the stable
solution or, equivalently, the pressure is maximal.
The latter corresponds to the effective action 
\beq
\Gamma = \frac{1}{\vv} \mbox{\textbf{Tr}}\ln \SI - \frac{1}{\vv}\mbox{\textbf{Tr}}  \lb \mathbbm{1}
-\frac{Z_2}{\vv^2}\SI_0 S \rb + \Gamma_2\,,
\eeq
where the traces are over momentum, color, flavor and Dirac components. 
The last term denotes the two-particle irreducible interaction part.
In our truncation scheme it corresponds to the diagram shown in Fig.~\ref{feyn_gamma2}
and is given by
\bseq
\Gamma_2 = & \frac{g^2}{2} \frac{1}{\vv^4}\sum\limits_{p,p',q}
\Tr\left[\Gamma_\mu^{a,0} S(q,q')\Gamma^a_\nu(q',p')D_{\mu\nu}(k)S(p',p)\right],
\eseq
where we have written the momentum sums explicitly, so that
the trace is only over internal degrees of freedom. 

The variation of the effective action with respect to the dressed propagator,
$\frac{\delta\Gamma}{\delta S(p,p')} = 0$,
just leads to the quark DSE \eq{eq:qdse} with the selfenergy \eq{eq:qsig}.
In addition, the effective action must be stationary with respect to the wave number, 
$\frac{d\Gamma}{d Q} =0$.
Denoting the dressing function of the dressed propagator proportional to $ip_3\gamma_5\gamma_3$ by $e_5(p)$, in analogy to the dressing function $E_5$ of the inverse propagator of \eq{eq:SIQ}, 
this condition can be simplified to 
\beq
\label{eq:Qgap}
\frac{1}{\vv}\sum_p p_3 e_5(p) \overset{!}{=} 0.
\eeq
For a homogeneous quark propagator, we have $e_5(p)=0$, and this equation is fulfilled trivially. 
For inhomogeneous propagators, on the other hand, 
it yields the additional constraint we need for determining $Q$. For this purpose, we solve the quark DSE \eq{eq:qdse} for different but fixed values of $Q$ and evaluate \eq{eq:Qgap} with these solutions. The zero of the left-hand side of \eq{eq:Qgap} then gives us the value of $Q$ that extremizes the effective action.

\begin{figure}[t]
	\centering
	\includegraphics[width=7.6cm]{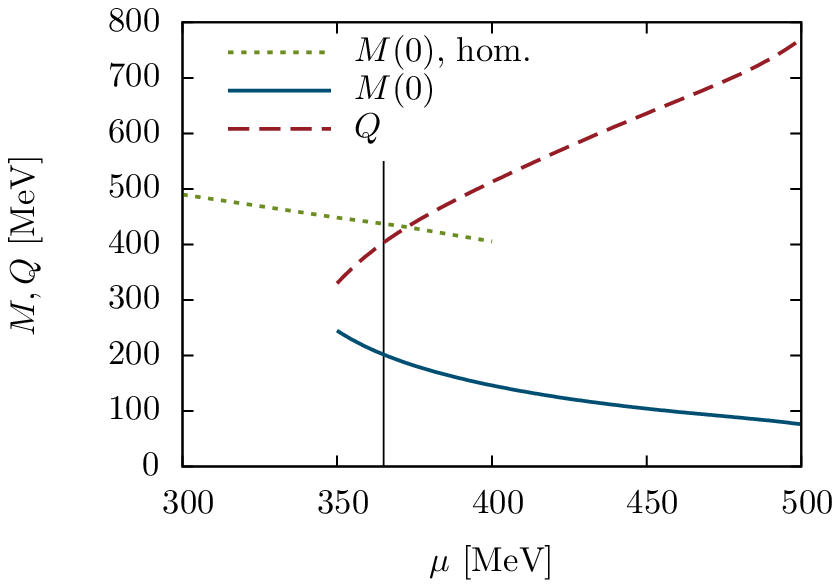}
	\centering
	\includegraphics[width=7.6cm]{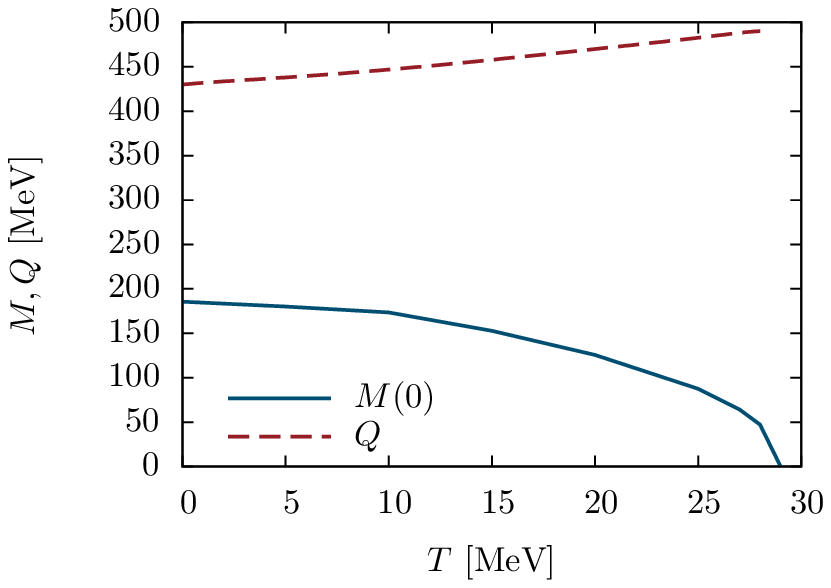}
\caption{Mass amplitude $M(0)$ and wave number $Q$ as functions of $\mu$ 
at $T=0$ (top) and of $T$ at $\mu=370$ MeV (bottom).
In the upper panel, the position of the first-order phase transition is indicated
by the thin vertical line.}
\label{fig:mQ_muT}
\end{figure}

We now  take the infinite-volume limit as specified in \eq{eq:infV} and
solve \eqs{eq:qdse} and (\ref{eq:Qgap}) numerically. 
Results for the mass amplitude $M(0) \equiv B(\vec 0,n=0)/C(\vec 0, n=0)$
and the wave number $Q$ are presented in  \fig{fig:mQ_muT}.
In the upper panel we show them for $T=0$ as functions of $\mu$.
At low chemical potential, we only find a homogeneous solution, whereas above 
$\mu = 350$~MeV, there is also an inhomogeneous solution.
Comparing the pressure, we find that the inhomogeneous phase
becomes favored above a critical chemical potential of about 365~MeV.
At this point a first-order phase transition takes place, 
where the wave number jumps from zero to a finite value,  
while the mass amplitude drops discontinuously, but remains nonzero. 
When we increase the chemical potential further, the latter continues to decrease,
but with decreasing slope, suggesting that the inhomogeneous phase survives 
up to arbitrarily high chemical potentials. 
In fact, within numerical accuracy, we never find the chirally restored phase to be
favored at $T=0$.

The lower panel shows the same quantities at $\mu=370$ MeV as functions of temperature. 
At this chemical potential the system is inhomogeneous at low temperatures, as evident from
the nonvanishing $Q$. With increasing temperature, $Q$ increases further, while the mass 
continuously decreases and becomes zero at about $T = 29$~MeV, 
where a second-order phase transition to the restored phase takes place.

\begin{figure}
	\centering
	\includegraphics[width=7.6cm]{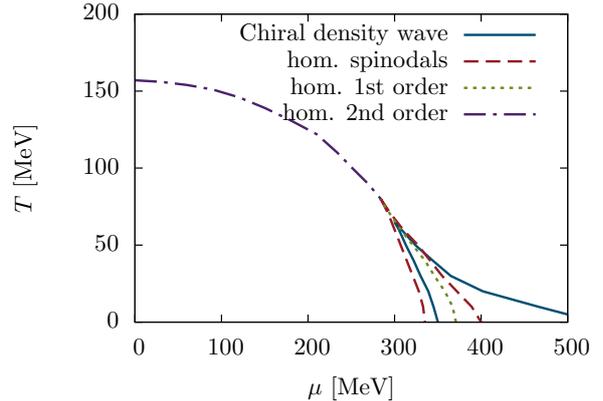}
\caption{Phase diagram in the $\mu-T$ plane. 
The blue solid line indicates the boundaries of the region where the inhomogeneous solution exists.}
\label{fig:pdQ}
\end{figure}

Collecting the results from different temperatures and chemical potentials we obtain the 
phase diagram displayed in \fig{fig:pdQ}. 
At high temperature there is a second-order phase transition between the homogeneous
chirally broken phase and the chirally restored phase (purple dash-dotted line), which becomes
first order at low temperatures (green dotted line) when the analysis is restricted to
homogeneous phases. The corresponding spinodals are indicated by the red dashed lines.

The limits of the region where we find an inhomogeneous solution with our CDW ansatz
are marked by the blue solid lines. 
It can be deduced from the behavior of $\frac{d\Gamma}{dQ}(Q)$ that 
in this region the inhomogeneous phase is always favored over the restored phase. 
The restored phase is eventually reached in a second-order phase transition.  
Moreover, from the fact that the homogeneous chirally broken phase is energetically degenerate
with the restored phase along the green dotted line, we conclude that the phase transition from 
the homogeneous to the inhomogeneous chirally broken phase must be to the left of this line.
As already seen for $T=0$, this phase transition is first order. 
Hence, the phase boundary must be somewhere between the left solid and the dotted line.
To locate it more precisely,  we have to compare the pressure of the two solutions,
which is numerically quite demanding. Within numerical precision we find that at $T=0$ 
the critical chemical potential is about 10~MeV lower than in the homogeneous case. 

Our most important result is that, within numerical resolution, the inhomogeneous phase covers the 
first-order phase boundary of the homogeneous case completely. 
Moreover, the point where the inhomogeneous phase and the two homogeneous phases meet 
seems to coincide with the homogeneous critical point.  
In this respect, \fig{fig:pdQ} has great similarities with the phase diagram in the 
Nambu--Jona-Lasinio (NJL) model~ \cite{Nickel:2009wj}.
A qualitative difference is that in the NJL model the inhomogeneous phase ends at some upper
critical chemical potential, whereas in the present case it seems to extent to arbitrarily high $\mu$
at $T=0$. 

On the other hand, at high densities inhomogeneous chiral symmetry breaking should become 
disfavored against quark pairing (color superconductivity)~\cite{Shuster:1999tn}, 
which we have neglected here.
We have recently studied color superconductivity in a similar framework~\cite{Muller:2013pya}
and it should be a feasible task to extend the present analysis in this direction. 
Additionally it needs to be checked if the results of this work are robust under the improvement of 
the truncation.
The consideration of more complicated inhomogeneous structures than the 
CDW ansatz would be interesting but extremely difficult as the propagator can no longer be inverted analytically. 

\section*{Acknowledgements} 
We thank Stefano Carignano for valuable discussions.
D.M. was supported by BMBF under contract 06DA9047I and by the Helmholtz Graduate School for Hadron and Ion Research.
M.B. and J.W. acknowledge partial support by the Helmholtz International
Center for FAIR and by the Helmholtz Institute EMMI.


\bibliography{literature}

\begin{thebibliography}{10}

\bibitem{BraunMunzinger:2009zz}
P.~Braun-Munzinger and J.~Wambach,
\newblock Rev.Mod.Phys. {\bf 81}, 1031 (2009).

\bibitem{Fukushima:2010bq}
K.~Fukushima and T.~Hatsuda,
\newblock Rept.Prog.Phys. {\bf 74}, 014001 (2011),
  \href{http://arxiv.org/abs/1005.4814}{{\ttfamily arXiv:1005.4814 [hep-ph]}}.

\bibitem{Borsanyi:2010bp}
Wuppertal-Budapest Collaboration, S.~Borsanyi {\em et~al.},
\newblock JHEP {\bf 1009}, 073 (2010),
  \href{http://arxiv.org/abs/1005.3508}{{\ttfamily arXiv:1005.3508 [hep-lat]}}.

\bibitem{Bazavov:2011nk}
A.~Bazavov {\em et~al.},
\newblock Phys.Rev. {\bf D85}, 054503 (2012),
  \href{http://arxiv.org/abs/1111.1710}{{\ttfamily arXiv:1111.1710 [hep-lat]}}.

\bibitem{Asakawa:1989bq}
M.~Asakawa and K.~Yazaki,
\newblock Nucl.Phys. {\bf A504}, 668 (1989).

\bibitem{Fischer:2009gk}
C.~S. Fischer and J.~A. M{\"u}ller,
\newblock Phys.Rev. {\bf D80}, 074029 (2009),
  \href{http://arxiv.org/abs/0908.0007}{{\ttfamily arXiv:0908.0007 [hep-ph]}}.

\bibitem{Fischer:2010fx}
C.~S. Fischer, A.~Maas, and J.~A. M{\"u}ller,
\newblock Eur.Phys.J. {\bf C68}, 165 (2010),
  \href{http://arxiv.org/abs/1003.1960}{{\ttfamily arXiv:1003.1960 [hep-ph]}}.

\bibitem{Fischer:2011mz}
C.~S. Fischer, J.~L{\"u}cker, and J.~A. M{\"u}ller,
\newblock Phys.Lett. {\bf B702}, 438 (2011),
  \href{http://arxiv.org/abs/1104.1564}{{\ttfamily arXiv:1104.1564 [hep-ph]}}.

\bibitem{Fischer:2012vc}
C.~S. Fischer and J.~L{\"u}cker,
\newblock Phys.Lett. {\bf B718}, 1036 (2013),
  \href{http://arxiv.org/abs/1206.5191}{{\ttfamily arXiv:1206.5191 [hep-ph]}}.

\bibitem{Broniowski:2011ef}
W.~Broniowski,
\newblock Acta Phys.Polon.Supp. {\bf 5}, 631 (2012),
  \href{http://arxiv.org/abs/1110.4063}{{\ttfamily arXiv:1110.4063 [nucl-th]}}.

\bibitem{Migdal:1973zm}
A.~Migdal,
\newblock Phys.Rev.Lett. {\bf 31}, 257 (1973).

\bibitem{Dautry:1979bk}
F.~Dautry and E.~Nyman,
\newblock Nucl.Phys. {\bf A319}, 323 (1979).

\bibitem{Broniowski:1990dy}
W.~Broniowski, A.~Kotlorz, and M.~Kutschera,
\newblock Acta Phys.Polon. {\bf B22}, 145 (1991).

\bibitem{Nakano:2004cd}
E.~Nakano and T.~Tatsumi,
\newblock Phys.Rev. {\bf D71}, 114006 (2005),
  \href{http://arxiv.org/abs/hep-ph/0411350}{{\ttfamily arXiv:hep-ph/0411350
  [hep-ph]}}.

\bibitem{Deryagin:1992rw}
D.~Deryagin, D.~Y. Grigoriev, and V.~Rubakov,
\newblock Int.J.Mod.Phys. {\bf A7}, 659 (1992).

\bibitem{Shuster:1999tn}
E.~Shuster and D.~Son,
\newblock Nucl.Phys. {\bf B573}, 434 (2000),
  \href{http://arxiv.org/abs/hep-ph/9905448}{{\ttfamily arXiv:hep-ph/9905448
  [hep-ph]}}.

\bibitem{Nickel:2009ke}
D.~Nickel,
\newblock Phys.Rev.Lett. {\bf 103}, 072301 (2009),
  \href{http://arxiv.org/abs/0902.1778}{{\ttfamily arXiv:0902.1778 [hep-ph]}}.

\bibitem{Nickel:2009wj}
D.~Nickel,
\newblock Phys.Rev. {\bf D80}, 074025 (2009),
  \href{http://arxiv.org/abs/0906.5295}{{\ttfamily arXiv:0906.5295 [hep-ph]}}.

\bibitem{Muller:2013pya}
D.~M{\"u}ller, M.~Buballa, and J.~Wambach,
\newblock Eur.Phys.J. {\bf A49}, 96 (2013),
  \href{http://arxiv.org/abs/1303.2693}{{\ttfamily arXiv:1303.2693 [hep-ph]}}.

\end{thebibliography}
\bibliographystyle{h-physrev-arxiv}

\end{document}